\documentclass[12pt,letterpaper]{article}
\usepackage[utf8]{inputenc}
\usepackage{amsmath}
\usepackage{mathptmx, courier,textcomp}
\usepackage{amsfonts}
\usepackage{amssymb}
\usepackage{graphicx}
\usepackage{cite} 
\usepackage{ulem}
\usepackage{caption}
\usepackage{epstopdf}
\usepackage[top=1in,bottom=1in,left=1in,right=1in]{geometry} 
\usepackage{float}
\usepackage{hyperref}
\usepackage{booktabs}
\usepackage[export]{adjustbox}
\usepackage{float}
\usepackage{multirow}
\usepackage{pifont}
\usepackage[ ]{authblk}
\usepackage{chemarrow}
\usepackage{comment}
\usepackage{makecell}
\usepackage{physics}
\UseRawInputEncoding
\hypersetup{
    colorlinks=true,
    linkcolor=blue,
    filecolor=magenta,      
    urlcolor=blue,
    pdftitle={Overleaf Example},
    pdfpagemode=FullScreen,
}
\begin{document}

%
%
\title{Tailoring the Frequency-Dependent Optical Response of Hematite through Mono- and Co-Doping: A First-Principles Study}
 
%
%

%
%
\author{Abdul Ahad Mamun}
\author[1,*]{Muhammad Anisuzzaman Talukder}
\affil{\small{Department of Electrical and Electronic Engineering\\

Bangladesh University of Engineering and Technology\\

Dhaka 1205, Bangladesh\\}}
\affil[ ]{\small{\it{$^*$anis@eee.buet.ac.bd}}}
%
%

\date{}
\maketitle

\newcommand{\OO}{O$_2$~}
\newcommand{\HH}{H$_2$~}
\sloppy
%
%
\begin{abstract}

Understanding the effects of doping on the crystal structure and optical properties of semiconductor materials is crucial for advancing next-generation semiconductor and photonic technologies. Although various studies have focused on doped hematite ($\alpha$-Fe$_2$O$_3$), the relationship between dynamical stability and optical properties remains insufficiently explored, limiting the rational design of functional materials. This study presents a comprehensive first-principles investigation that simultaneously evaluates the phonon dispersion characteristics and frequency-dependent optical response of B-doped, Y-doped, and (B, Y)-co-doped  $\alpha$-Fe$_2$O$_3$, providing deeper insights into the underlying mechanisms. We examined the finite-temperature vibrational properties, dielectric function, and optical characteristics to comprehend the lattice dynamics and light-matter interactions under electromagnetic radiation. Vibrational thermodynamics reveal that pristine and Y-doped $\alpha$-Fe$_2$O$_3$ maintain dynamic stability, while B-doped $\alpha$-Fe$_2$O$_3$ exhibits imaginary phonon modes indicating lattice instability due to distortions in the Fe--O framework. Notably, Y co-doping with B helps suppress these soft modes, restoring structural stability through lattice relaxation and improved interatomic forces. B doping enhances low-energy absorption by introducing additional states in the valence band, while Y doping alters orbital hybridization, leading to a broader dispersion. In the optical regime, doped $\alpha$-Fe$_2$O$_3$ displays dominant interband transitions below $2$ eV and strong absorption between $1.80$ eV and $4$ eV. The (B, Y) co-doping combines the low-energy benefits with an improved optical response profile. In summary, doping significantly enhances lattice vibrations, light-matter interactions, and optical responses, providing an effective strategy for tailoring $\alpha$-Fe$_2$O$_3$ for diverse applications in photoactive, optoelectronic, and photonic technologies.

\end{abstract}
%
%

%
%

%
%
\section{Introduction}

Hematite ($\alpha$-Fe$_2$O$_3$) is a thermodynamically stable semiconducting iron oxide, recognized as a prototypical transition-metal oxide \cite{wan2023advanced}. It is extensively studied in modern optoelectronics, photonics, and solar-driven energy conversion due to its abundance, stability, and ability to interact with visible and near-infrared light \cite{mamun2025advancing,alziyadi2025hematite}. Hematite has emerged as a key photoelectrode material for photoelectrochemical (PEC) water splitting, offering a theoretical solar-to-hydrogen efficiency of $\sim 15$\%, eventually making it one of the most promising candidates for sustainable hydrogen production \cite{mamun2025advancing,sivula2011solar}. Beyond PEC systems, hematite is becoming important in various optoelectronic devices, such as photodetectors, gas sensors, and spin-based systems \cite{chakraborty2022alpha,das2022mos2}. It is also utilized in advanced optical technologies, including wavelength-selective coatings and photonic modulation platforms \cite{kubacka2014large}. The performance of these systems is fundamentally governed by the material's frequency-dependent optical response, which affects light absorption efficiency, refractive index dispersion, dielectric polarization, optical conductivity ($\sigma_{\rm op}$), reflectivity ($R_{\rm op}$), and penetration depth ($\delta_{\rm op}$) \cite{kronawitter2011doped,yousaf2019tuning}. These optical properties are closely related to interband electronic transitions within the UV-visible range, which play a critical role in photon harvesting and the efficiency of electromagnetic energy conversion \cite{piccinin2019band}. Recent advancements in materials engineering have indicated that tailoring the optical response of hematite is a foremost strategy for optimizing its multifunctional performance across various optoelectronic and energy-related applications \cite{xue2020review}.

Despite its wide range of potential applications, pristine $\alpha$-Fe$_2$O$_3$ has inherent optical limitations that reduce its effectiveness in high-performance optical and optoelectronic devices \cite{mamun2025improved,yin2017experimental}. It has a relatively large bandgap energy ($E_g$), low charge carrier mobility, and significant electron-hole recombination, all of which reduce optical efficiency and decrease the conductivity \cite{xue2020review}. Additionally, the low dielectric response of pristine $\alpha$-Fe$_2$O$_3$ often leads to suboptimal performance in light confinement \cite{lunt2013dielectric}. This limitation impedes polarization-dependent optical manipulation, which is essential for meta-surfaces and polarization-sensitive photonic devices. As an optical material, pristine $\alpha$-Fe$_2$O$_3$ shows limited tunability of its refractive index ($n$) and only modest dielectric constant ($\epsilon_{\rm re,0}$) \cite{bhattacharjee2021field,kumar2015controlling}. These characteristics restrict its use in multilayer interference coatings, waveguides, and photonic crystal structures. Moreover, controlling the extinction coefficient ($k$) of pristine $\alpha$-Fe$_2$O$_3$ near the absorption edge is challenging, leading to weak spectral selectivity and reduced efficiency in wavelength filtering and optical switching applications \cite{kumar2015controlling}.

In PEC water-splitting systems, optical limitations can lead to inefficient light harvesting \cite{wang2017hematite,mamun2025improved}. The inadequate absorption of longer visible wavelengths decreases the depth of photon utilization, limiting the generation of photocarriers under solar illumination. The moderate $\sigma_{\rm op}$ and the slow onset of interband transitions impede effective photon-to-charge conversion, ultimately reducing the overall efficiency of solar-to-hydrogen conversion \cite{yin2017experimental}. Furthermore, advanced photonic and nano-optical devices face additional challenges due to weak electromagnetic field enhancement and limited plasmonic compatibility, as pristine $\alpha$-Fe$_2$O$_3$ does not inherently exhibit strong free-carrier-driven optical resonances \cite{thimsen2011influence,su2024steering}.

Several strategies have been employed to optimize the optical and electronic properties of $\alpha$-Fe$_2$O$_3$, including nanostructuring, heterostructuring, surface modification, and elemental doping \cite{tofanello2020strategies,park2023recent,souza2025revisiting,wan2023advanced}. These methods are widely used to enhance light absorption, charge transport, and optical response, thereby improving performance in PEC, optoelectronic, and optical material applications \cite{wan2023advanced}. In PEC water splitting, Sivula et al.~utilized nanostructuring and interface engineering as primary strategies to boost charge separation and increase the solar-to-hydrogen efficiency of $\alpha$-Fe$_2$O$_3$ photoelectrodes \cite{sivula2011solar}. By controlling the microstructure and porosity, polycrystalline $\alpha$-Fe$_2$O$_3$ thin films exhibited tunable refractive index, extinction coefficient, and absorption coefficient \cite{glasscock2008structural}. Moreover, the controlled growth of $\alpha$-Fe$_2$O$_3$ films can alter the bandgap energy and refractive index, suggesting potential applications in optical coatings and light management systems \cite{glasscock2008structural,sarkar2024structure}. Similarly, Souza et al.~conducted an ellipsometric study on Sn-doped $\alpha$-Fe$_2$O$_3$ by varying the annealing temperature to explore structural evolution and its effects on the optical response. In optoelectronic applications, Maz{\'o}n-Montijo et al.~reported that nanostructured $\alpha$-Fe$_2$O$_3$ thin films exhibit measurable photoconductivity and a visible-light photoresponse arising from photon-induced electron–hole generation followed by defect-mediated carrier transport, indicating effective optical sensing behavior \cite{mazon2020role}. These studies demonstrate that $\alpha$-Fe$_2$O$_3$ can perform as a visible-light-responsive optoelectronic material, although its performance is significantly influenced by defect engineering and microstructural control.

Recently, researchers have made significant progress in modifying $\alpha$-Fe$_2$O$_3$ through compositional engineering and various doping strategies, resulting in distinct optical responses \cite{elouafi2020effects,suman2020zn,irshad2024exploring}. Mono-doping typically alters the optical absorption and interband transition features, subsequently affecting the characteristics at the band edge \cite{shelton2022polaronic}. In contrast, co-doping introduces more complex changes in dielectric behavior and the dispersion of the refractive index, driven by interactions between defects and the host atoms \cite{jha2025beyond}. These changes are often reflected in frequency-dependent optical functions, such as complex permittivity ($\epsilon$), $\sigma_{\rm op}$, and $k$ \cite{irshad2024exploring,kumar2015controlling}. Additionally, the modified $\alpha$-Fe$_2$O$_3$ may exhibit phonon-sensitive structural stability, where lattice vibrations can influence optical transitions and light-matter interactions via electron-phonon coupling \cite{shelton2022polaronic}. Within this corresponding framework of optical and vibrational properties, first-principles approaches provide a comprehensive method for consistently evaluating both dynamic stability and frequency-dependent optical properties in doped $\alpha$-Fe$_2$O$_3$, enabling a deeper understanding of the optical functionality associated with mono- and co-doping strategies.

%

Our previous research established the defect energetics, electronic structures, and charge transport properties of boron (B)-doped, yttrium (Y)-doped, and (B, Y) co-doped $\alpha$-Fe$_2$O$_3$ \cite{mamun2025improved,mamun2026re}. However, several important questions remain unresolved, particularly concerning the dynamic stability of the predicted doped structures and the impact of dopant-induced lattice modifications on their vibrational properties and frequency-dependent optical characteristics. To address these issues, we conduct a comprehensive analysis of phonon behavior, mode-resolved vibrational spectra, dielectric response, and optical tensor properties in this study. Specifically, we systematically investigate the dynamic structural stability and optical behaviors of pristine $\alpha$-Fe$_2$O$_3$, along with B-doped, Y-doped, and (B, Y) co-doped $\alpha$-Fe$_2$O$_3$, utilizing first-principles calculations in conjunction with the Phonopy method. 

We assessed the dynamical stability of each sample through phonon dispersion analysis, along with calculations of vibrational free energy ($F_{\rm vib}$) and entropy ($S_{\rm vib}$). Our findings reveal that B-doped hematite exhibits phonon instability. In contrast, both Y-doped and (B, Y) co-doped systems demonstrate dynamically stable behavior, described by entirely positive phonon modes. To examine the light-matter interaction in the doped $\alpha$-Fe$_2$O$_3$, we conducted a detailed analysis of the optical responses and calculated $E_{g}$. The parameter $\epsilon$ was determined to represent the frequency-dependent polarization response under electromagnetic radiation. 

In addition, we determined $n$ and $k$ to better understand light propagation, phase modulation, and absorption behavior within the material. The electric susceptibility ($\chi_E$) was used to quantify the strength of dielectric polarization under external fields. The parameter $\sigma_{\rm op}$ was computed to evaluate photon-induced charge transport and the efficiency of electronic excitation. The analysis of $R_{\rm op}$ was performed to assess the surface optical response, while we examined $\delta_{\rm op}$ to characterize the extent of light-matter interaction within the material. The results indicate that Y doping and (B, Y) co-doping significantly enhance both structural stability and optical performance compared to pristine and B-doped $\alpha$-Fe$_2$O$_3$. This improvement suggests superior photon-matter interaction and greater tunability of optical properties. Overall, our findings provide valuable insights into the relationship between doping-induced lattice dynamics and optical responses in $\alpha$-Fe$_2$O$_3$, offering potential applications in next-generation optoelectronic, photonic, and solar energy conversion technologies.

%
%

%
%
\section{Computational Methodology} 

We performed first-principles calculations within the framework of spin-polarized density functional theory (DFT) using the self-consistent ab initio approach implemented in the Quantum Espresso (QE) software \cite{giannozzi2009quantum}. To estimate the exchange-correlation functions, we employed the general gradient approximation (GGA) with the Perdew-Burke-Ernzerhof (PBE) model \cite{giannozzi2009quantum,giannozzi2017advanced}. The projected augmented wave (PAW) methodology was utilized to model the core electrons, while the valence electrons were described by Kohn-Sham (KS) functions, expanded in a plane-wave basis with a kinetic energy cutoff of 60 Ry \cite{kresse1999ultrasoft}. For the charge density, we set the cutoff kinetic energy to be approximately eight times greater than that of the wave functions. To ensure electronic convergence, we applied the Marzari-Vanderbilt smearing scheme with a width of 0.01 Ry. To accurately represent the localized d-orbitals of transition metal atoms, we employed the DFT$+$U framework developed by Dudarev et al.~\cite{dudarev1998electron}. The effective Hubbard correction term was set to $4.30$, as symmetrically determined by Mosey et al.~\cite{mosey2008rotationally}. Long-range dispersion interactions were incorporated using the DFT-D3 correction developed by Grimme et al.~\cite{grimme2010consistent}. Brillouin-zone integrations were carried out with Monkhorst-Pack $k$-point meshes of $7 \times 7 \times 7$ \cite{monkhorst1976special}. We fully optimized the lattice parameters and atomic positions using the Broyden-Fletcher-Goldfarb-Shanno (BFGS) algorithm. This optimization process continued until the total energy and atomic forces converged to $10^{-6}$ eV and $10^{-3}$ eV/\AA, respectively.

The atomic structures of both pristine and doped hematite systems are shown in Fig.~S1. We examined all possible substitutional positions in the host lattice to identify configurations that are both energetically and structurally stable. For single-dopant configurations using B and Y, the designed doping concentration was $4.719 \times 10^{20}$ cm$^{-3}$, which corresponds to a single-atom substitution in the unit cell crystal structure. For co-doped configurations with (B, Y), the concentration was $9.439 \times 10^{20}$ cm$^{-3}$. This consistent approach enables a reliable comparison of the structural, vibrational, and optical properties across all investigated cases.

The dynamical stability of the optimized structures was assessed through phonon dispersion calculations using the finite-displacement method in the Phonopy package \cite{togo2023implementation,togo2023first}. Supercells were generated from the relaxed structures, and harmonic interatomic force constants were derived from these supercells using an atomic displacement amplitude of 0.01 \AA. A $7\times 7\times 7$ $k$-point mesh was used for force calculations, while a denser $15\times 15\times 15$ $q$-point mesh was employed to compute the phonon dispersion relations and the phonon density of states (PDOS). $F_{\rm vib}$ and $S_{\rm vib}$ were calculated using the harmonic approximation to assess thermodynamic stability.  

To elucidate the effects of doping on the interaction between electromagnetic radiation and $\alpha$-Fe$_2$O$_3$, we conducted an in-depth analysis of the frequency-dependent optical response. These optical properties provide essential insights into photon absorption, dielectric polarization, electronic excitation, and light propagation. To investigate these optical characteristics, we calculated the $\epsilon$, within the framework of linear-response theory, applying the independent-particle random phase approximation (RPA) based on the DFT electronic structure. The $n$, $k$, $\chi_E$, $\sigma_{\rm op}$, $R_{\rm op}$, and $\delta_{\rm op}$ can be obtained from the real ($\epsilon_{\rm re}$) and imaginary ($\epsilon_{\rm im}$) parts of $\epsilon$ using the following equations \cite{fox2010optical,peter2010fundamentals,born2013principles}
\begin{subequations}
\allowdisplaybreaks
\begin{align}
    & n(\omega) = \frac{1}{\sqrt{2}} \left[\sqrt{\epsilon_{\rm re}^2(\omega) + \epsilon_{\rm im}^2(\omega)} + \epsilon_{\rm re}(\omega) \right]^{1/2}, \\
    & k(\omega) = \frac{1}{\sqrt{2}} \left[ \sqrt{\epsilon_{\rm re}^2(\omega) + \epsilon_{\rm im}^2(\omega)} - \epsilon_{\rm re}(\omega) \right]^{1/2}, \\
    & \chi_E(\omega) = \epsilon(\omega) - 1, \\
    & \sigma_{\rm op}(\omega) = i \omega \epsilon_0 \chi_E(\omega), \\
    & R_{\rm op}(\omega) = \frac{[ n(\omega) - 1 ]^2 + k^2(\omega)}{[ n(\omega) + 1 ]^2 + k^2(\omega)},\\
    & \delta_{\rm op}(\omega) = \frac{c}{2\omega k(\omega)},
    \label{Eq2.37}
\end{align}
\end{subequations}
where $\omega$ is the angular frequency, which can be expressed as $\omega = (2\pi c)/{\lambda}$. Here, $c$ and $\lambda$ are the speed of light and optical wavelength, respectively.

%
%

%
\begin{figure}[htb]
    \centering
    \includegraphics[width =0.89\linewidth]{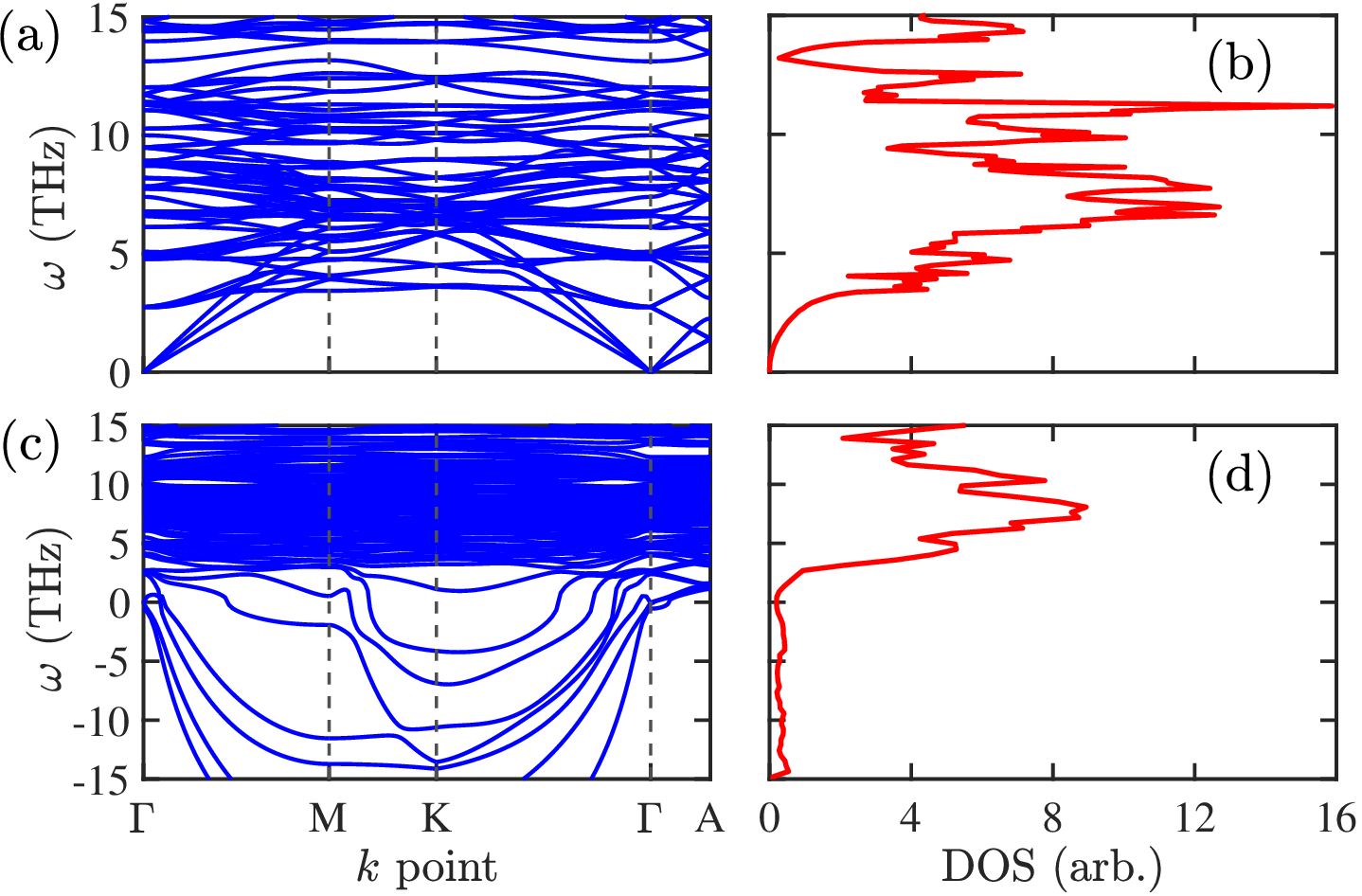}
    \caption{(a) Phonon band structure and (b) phonon density of states (PDOS) of pristine $\alpha$-Fe$_2$O$_3$. (c) Phonon band structure and (d) phonon density of states (PDOS) of B-doped $\alpha$-Fe$_2$O$_3$.}
    \label{Fig1}
\end{figure}
%
%

%
%
\section{Results and Analysis}
\subsection{Dynamically Stable Crystal Structure}

Phonon dispersion relations and the PDOS are essential parameters for understanding lattice dynamics in crystalline solids \cite{passler2007basic}. They provide important insights into structural stability, interatomic force interactions, thermal transport properties, and the mechanisms of electron-phonon coupling. The criterion for dynamical stability is the absence of imaginary phonon frequencies across the entire Brillouin zone \cite{alarco2018phonon}. The presence of imaginary modes indicates a dynamical instability in the crystal structure, which often suggests a tendency toward a disordered phase transition due to soft phonon modes. Furthermore, the PDOS reveals the distribution of vibrational modes across the frequency spectrum and the nature of chemical bonding. Typically, low-frequency modes are associated with heavy atoms and weaker bonding, while high-frequency modes are related to strong covalent bonding and lighter atomic masses. Additionally, phonon-related properties are crucial in determining thermal characteristics, carrier relaxation dynamics, and phonon-assisted mechanisms, which impact the functional performance of semiconducting and photoactive materials \cite{alarco2018phonon}.

Figures 1(a, b) and (c, d) illustrate the phonon band structures and the corresponding PDOS for pristine and B-doped $\alpha$-Fe$_2$O$_3$, respectively. The phonon band structure of pristine $\alpha$-Fe$_2$O$_3$ shows positive frequencies throughout the Brillouin zone, confirming its dynamical stability. This finding is consistent with the experimentally observed stability of the hematite phase \cite{padilha2019theoretical}. The substantial optical modes appear at frequencies of up to 15 THz, highlighting the complex nature of the hematite lattice and the rich vibrational interactions within the Fe--O network. The corresponding PDOS reveals several pronounced peaks between 4 and 12 THz, indicating high concentrations of vibrational states associated with collective lattice vibrations. 

Conversely, B-doped $\alpha$-Fe$_2$O$_3$ exhibits several phonon branches with imaginary frequencies in certain directions within reciprocal space, indicating the presence of unstable vibrational modes. The persistence of these soft modes across various doping arrangements suggests that B doping introduces significant lattice distortions, destabilizing the hematite framework. These findings imply that B-doped $\alpha$-Fe$_2$O$_3$ may spontaneously undergo structural reconstruction toward a different phase or local atomic arrangement. Moreover, the PDOS of B-doped $\alpha$-Fe$_2$O$_3$ shows a redistribution of vibrational states, particularly in the low-frequency region, reflecting changes in local force constants and bonding interactions due to B incorporation. 

%
\begin{figure}[htb]
    \centering
    \includegraphics[width =0.89\linewidth]{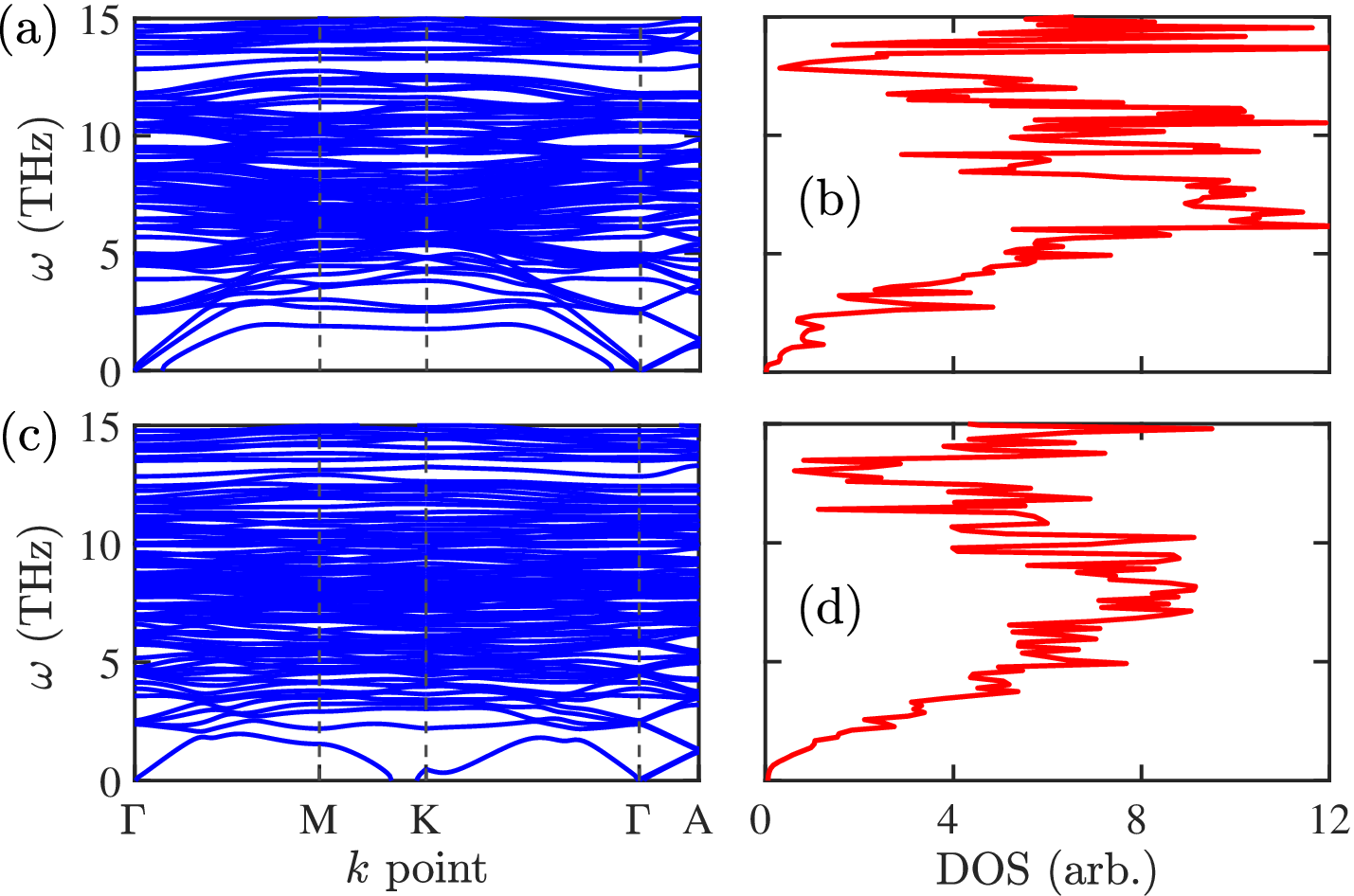}
    \caption{(a) Phonon band structure and (b) phonon density of states (PDOS) of Y-doped $\alpha$-Fe$_2$O$_3$. (c) Phonon band structure and (d) phonon density of states (PDOS) of (B, Y)-co-doped $\alpha$-Fe$_2$O$_3$.}
    \label{Fig2}
\end{figure}
%
%

Figures 2(a, b) and (c, d) display the calculated phonon band structures and the corresponding PDOS for Y-doped $\alpha$-Fe$_2$O$_3$ and (B, Y) co-doped $\alpha$-Fe$_2$O$_3$, respectively. In both cases, all phonon modes exhibit positive frequencies throughout the entire Brillouin zone. This observation confirms their dynamical stability, implying that the optimized structures correspond to local minima on the potential-energy surface. The acoustic branches begin at zero frequency at the $\Gamma$ point and evolve smoothly toward the zone boundaries, satisfying the translational invariance condition of the crystal lattice. Numerous optical modes are observed within the frequency range of 3 to 15 THz, reflecting complex vibrational modes of Fe--O, Y--O, and doping-induced interactions. The PDOS spectra reveal broad vibrational distributions with several distinct peaks in the intermediate and high-frequency regions, demonstrating significant contributions from O-related vibrations and bond-stretching modes. Notably, there is no photonic band gap in the doped $\alpha$-Fe$_2$O$_3$, which allows the continuous vibrational connectivity between the acoustic and optical branches, thus facilitating phonon-mediated energy transfer. The stable optical phonon modes extending up to 15 THz suggest robust lattice dynamics and favorable thermal stability under operating conditions. These characteristics are especially desirable for energy conversion and optoelectronic applications, where structural stability is essential for efficient charge transfer, improved carrier relaxation, and long-term operational durability.

A noteworthy finding from the phonon dispersion analysis is the opposite dynamical behavior between B-doped $\alpha$-Fe$_2$O$_3$ and (B, Y) co-doped $\alpha$-Fe$_2$O$_3$. While B-doped $\alpha$-Fe$_2$O$_3$ exhibits lattice instability, the incorporation of Y effectively stabilizes the crystal structure of (B, Y)-co-doped $\alpha$-Fe$_2$O$_3$. Y doping helps balance the local structural distortions caused by B doping. The relatively small atomic size and distinct bonding characteristics of B disrupt the Fe--O framework, resulting in unfavorable interatomic force constants and the emergence of soft phonon modes. Conversely, Y, which has a substantially larger ionic radius, forms strong Y--O bonds that facilitate local lattice relaxation and redistribute internal strain within the crystal. Consequently, the destabilizing effect observed in B-doped $\alpha$-Fe$_2$O$_3$ is mitigated, leading to the recovery of a dynamically stable structure in (B, Y)-co-doped $\alpha$-Fe$_2$O$_3$.

%
\begin{figure}[htbp]
    \centering
    \includegraphics[width =0.95\linewidth]{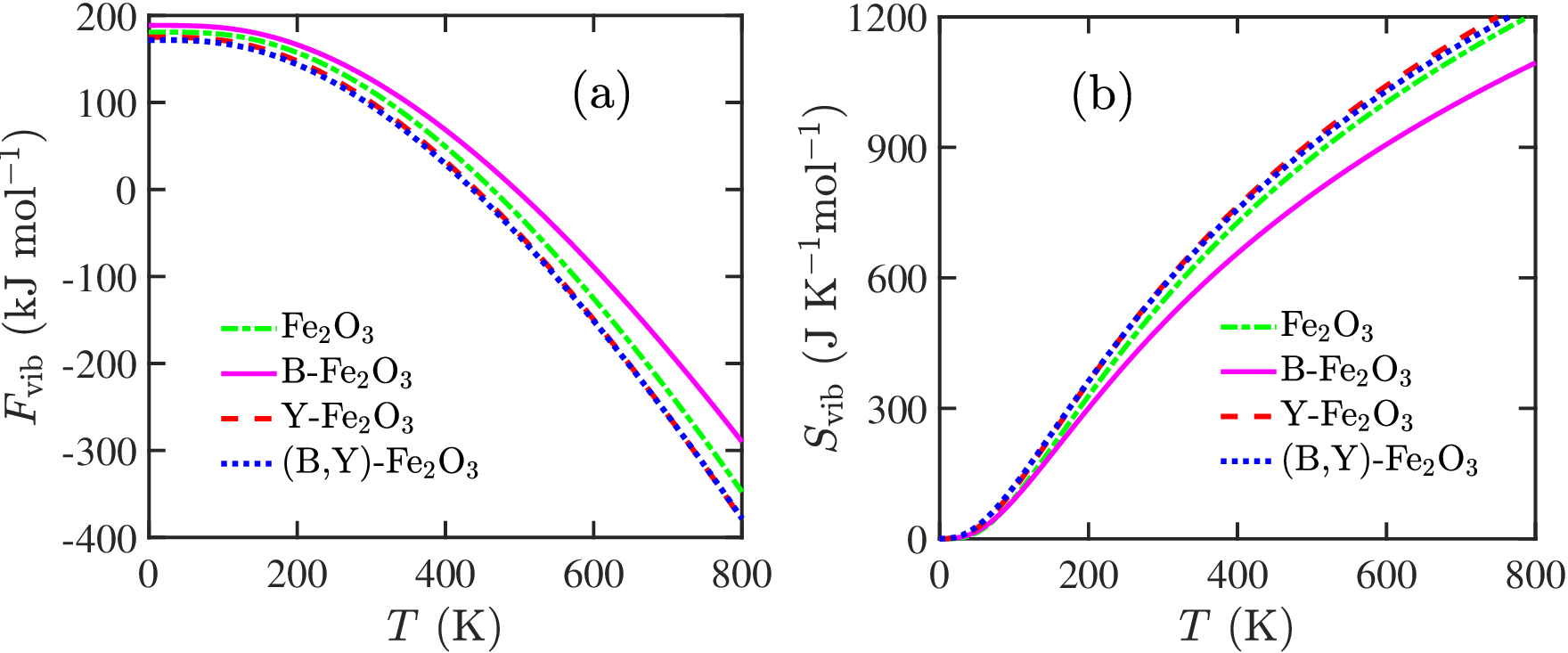}
    \caption{(a) Vibrational free energy, $F_{\rm vib}$, and (b) vibrational entropy, $S_{\rm vib}$, of pristine and doped $\alpha$-Fe$_2$O$_3$.}
    \label{Fig3}
\end{figure}
%
%

The temperature-dependent $F_{\rm vib}$ and $S_{\rm vib}$ provide valuable insights into the lattice dynamics and thermal stability of crystalline materials. These properties are directly derived from the phonon spectrum and reflect the total free energy of systems due to lattice vibrations. Specifically, $F_{\rm vib}$ describes the phononic contribution to the overall thermodynamic potential, while $S_{\rm vib}$ measures the extent of vibrational disorder caused by thermally populated phonon states \cite{zhao2024temperature}.

Figure 3 illustrates the temperature-dependent $F_{\rm vib}$ and $S_{\rm vib}$ for pristine, B-doped, Y-doped, and (B, Y)-co-doped $\alpha$-Fe$_2$O$_3$. As shown in Figure 3(a), $F_{\rm vib}$ decreases monotonically with increasing temperature ($T$) in all investigated cases due to the presence of higher-energy phonon states. This behavior enhances the vibrational contribution to free energy and reduces the overall thermodynamic potential of the crystal structures. At a room temperature of $T = 300$ K, $F_{\rm vib}$ is at 113.2 kJ mol$^{-1}$ for pristine $\alpha$-Fe$_2$O$_3$, 126.4 kJ mol$^{-1}$ for B-doped $\alpha$-Fe$_2$O$_3$, 96.28 kJ mol$^{-1}$ for Y-doped $\alpha$-Fe$_2$O$_3$, and 100.2 kJ mol$^{-1}$ for (B, Y)-co-doped $\alpha$-Fe$_2$O$_3$. As $T$ increases to $800$ K, $F_{\rm vib}$ decreases significantly, reaching values between approximately $-300$ and $-380$ kJ mol$^{-1}$. Among all systems, the (B, Y)-co-doped structure consistently exhibits the lowest $F_{\rm vib}$, indicating the most favorable thermodynamic contribution from vibrations.

Simultaneously, Fig.~3(b) shows that $S_{\rm vib}$ increases continuously with $T$ due to the enhanced thermal population of phonon modes and the corresponding increase in lattice disorder. At $T=300$ K, $S_{\rm vib}$ is approximately $500$--$580$ J K$^{-1}$mol$^{-1}$ for all systems, while at $T=800$ K, it approaches $1100$--$1200$ J K$^{-1}$mol$^{-1}$. The Y-doped and (B, Y)-co-doped structures display slightly higher $S_{\rm vib}$ than pristine $\alpha$-Fe$_2$O$_3$, indicating a greater number of thermally accessible vibrational states and enhanced vibrational flexibility of the lattice. This combination of dynamic stability, favorable $F_{\rm vib}$, and increased $S_{\rm vib}$ suggests that Y-doped and (B, Y)-co-doped $\alpha$-Fe$_2$O$_3$ are thermodynamically robust over a broad temperature range, ensuring long-term thermal stability and superior resistance to temperature-induced structural degradation, which is essential for reliable device performance.

\subsection{Optical Properties}

\subsubsection{Complex Permittivity and Optical Bandgap Energy}
The complex dielectric function provides important insights into the polarization behavior and optical excitation processes of materials \cite{fox2010optical}. Specifically, the $\epsilon_{\rm re}$ represents a material's ability to store electrical energy through polarization. In contrast, the $\epsilon_{\rm im}$ reflects the energy dissipation associated with photon-induced electronic transitions. These components are closely related to dielectric screening, charge-carrier dynamics, refractive properties, and optical absorption, making them fundamental parameters for assessing the electronic and photonic performance of semiconductor materials.

%
\begin{figure}[htbp]
    \centering
    \includegraphics[width =0.83\linewidth]{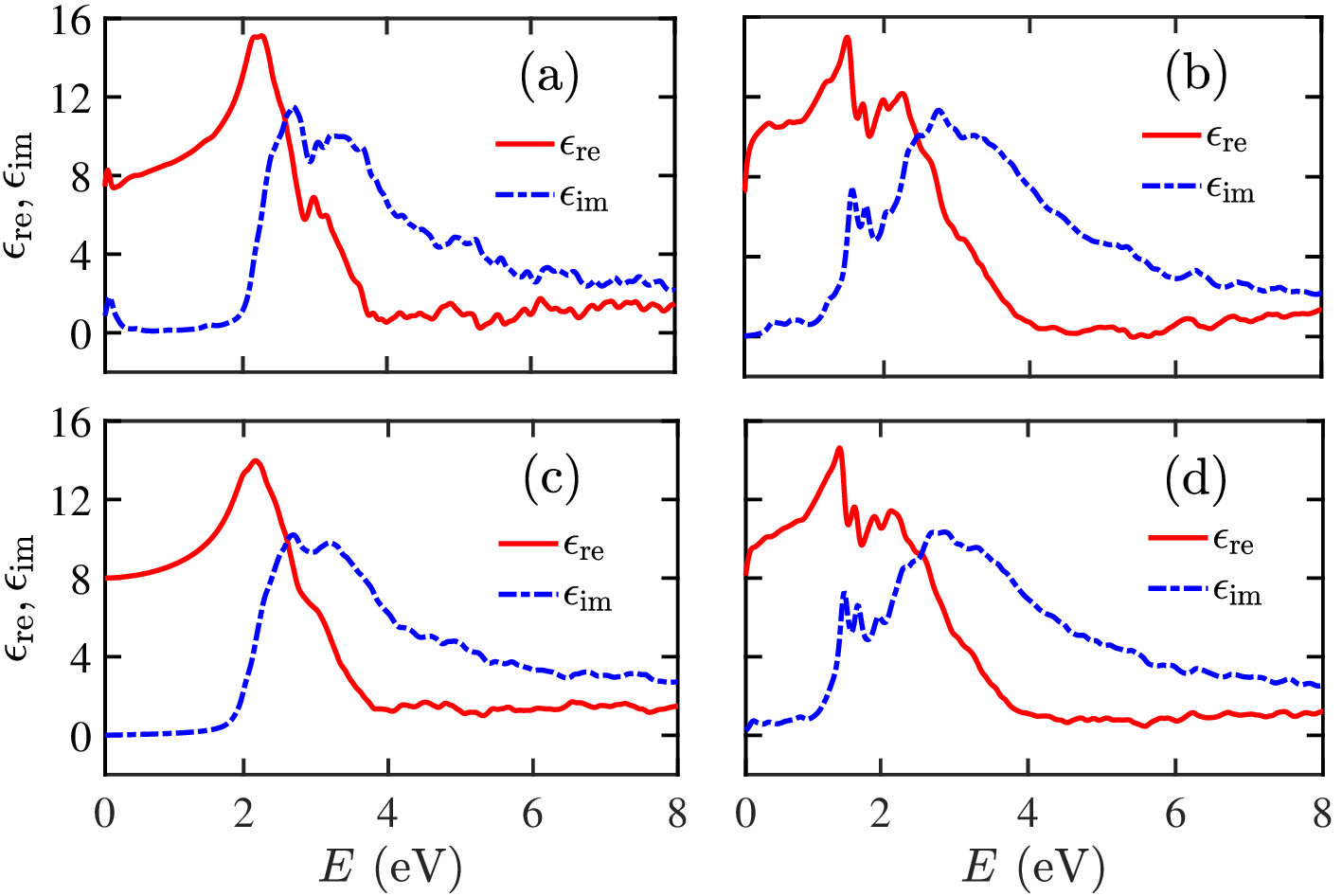}
    \caption{Real $(\epsilon_{\rm re})$, and imaginary parts $(\epsilon_{\rm im})$ of complex permittivity $(\epsilon)$ as a function of photon energy ($E$) for (a) pristine $\alpha$-Fe$_2$O$_3$, (b) B-doped $\alpha$-Fe$_2$O$_3$, (c) Y-doped $\alpha$-Fe$_2$O$_3$, and (d) (B, Y)-co-doped $\alpha$-Fe$_2$O$_3$.}
    \label{Fig4}
\end{figure}
%
%

Figure 4 illustrates $\epsilon_{\rm re}$ and $\epsilon_{\rm im}$ components of $\epsilon$ as functions of photon energy ($E$) for both pristine and doped $\alpha$-Fe$_2$O$_3$. For all doped samples, both $\epsilon_{\rm re}$ and $\epsilon_{\rm im}$ exhibit significant features within the energy range of 2 to 5 eV, which correlate with strong interband transitions characteristic of the electronic structure. In pristine $\alpha$-Fe$_2$O$_3$, the first notable peak in $\epsilon_{\rm re}$ occurs near 2 eV, likely due to transitions from the 3d orbitals of Fe to the hybridized 3d orbitals of Fe and 2p orbitals of O. Additionally, a pronounced peak between 3 and 4 eV suggests deeper valence-to-conduction band excitations. The static dielectric constant of pristine $\alpha$-Fe$_2$O$_3$ is $\epsilon_{\rm re,0} = 7.809$. The $\epsilon_{\rm re}$ increases sharply, reaching a maximum value of nearly 15.0 at $E \approx 2.2$ eV, while the corresponding loss peak occurs at $E \approx 2.6$ eV with $\epsilon_{\rm im}$ of $\sim 11.5$. However, the relatively abrupt onset of absorption near 2.30 eV indicates that a significant portion of visible and infrared lights is not effectively utilized, limiting the overall spectral coverage despite the favorable polarization response.

In B-doped $\alpha$-Fe$_2$O$_3$, the static dielectric constant decreases to $\epsilon_{\rm re,0}$ of 7.294, as shown in Fig.~4(b). Although the maximum $\epsilon_{\rm re}$ remains close to $15.0$, it shifts toward lower photon energies of $\sim 1.7$ eV. The strongest dielectric loss occurs near 2.5 eV, with $\epsilon_{\rm im}$ of $\sim 10.5$. Importantly, $\epsilon_{\rm im}$ exhibits a distinct sub-bandgap absorption tail below 2 eV, indicating the formation of B-induced localized states within the bandgap. These localized states provide additional optical excitation pathways, allowing for the absorption of lower-energy photons. Meanwhile, the enhanced low-energy $\epsilon_{\rm re}$ improves dielectric screening and weakens electron-hole binding. Consequently, B doping broadens the spectral absorption window and strengthens light-matter interactions, though excessive concentrations of these localized states may lead to carrier trapping and non-radiative recombination channels.

In Figure 4(c), the peaks in both the $\epsilon_{\rm re}$ and $\epsilon_{\rm im}$ for Y-doped $\alpha$-Fe$_2$O$_3$ slightly shift to lower energies and show reduced intensity. These changes suggest alterations in the Fe--O bond hybridization and crystal field splitting due to the different ionic radii of Y$^{3+}$ and Fe$^{3+}$. This subtle lattice distortion reduces the formation of deep trap states while shifting the transition energies, thereby enhancing the light absorption coefficient. The static dielectric constant slightly increased at $\epsilon_{\rm re,0}=8.002$, while the maximum $\epsilon_{\rm re}$ decreases barely to $\approx 14.0$ near $E = 2.1$ eV. Similarly, the peak value of $\epsilon_{\rm im}$ is reduced to $\approx 10.0$ and shifts to $E \sim 2.8$ eV. This redistribution of spectral weight and strong dielectric screening reveals improved carrier delocalization and reduced trap sites, both of which are desirable for efficient electronic transport and a stable optical response.

The co-doping strategy involving B and Y leads to more complex and balanced modifications of the dielectric properties, as illustrated in Fig.~4(d). In (B, Y)-doped $\alpha$-Fe$_2$O$_3$, Y modifies the crystal symmetry and hybridization, while B introduces new states in the valence band. This interaction results in a redistribution of spectral weight in $\epsilon_{\rm im}$ and moderated peak magnitudes, suggesting a synergy between improved carrier mobility and extended absorption coefficient. The static dielectric constant is at $\epsilon_{\rm re,0} = 8.195$, exceeding that of both pristine and Y-doped $\alpha$-Fe$_2$O$_3$. Meanwhile, the maximum value of $\epsilon_{\rm re}$ reaches $\sim 14.5$ near $E=1.6$ to $1.8$ eV. Concurrently, the peak value of $\epsilon_{\rm im}$ decreases to $10.0$ around $E = 2.6$ to $2.8$ eV and maintains a noticeable low-energy absorption tail below $E = 2$ eV. Compared to pristine and B-doped $\alpha$-Fe$_2$O$_3$, the dielectric loss spectrum of (B, Y) co-doping becomes smoother and less intense, indicating a more homogeneous distribution of electronic states and reduced localization effects. To further understand the optical isotropy/anisotropy and electronic excitation properties, we present the optical tensor and electron energy loss spectroscopy (EELS) spectra in Figs.~S2 and S3. The components of the optical tensor along the $x$, $y$, and $z$ axes display nearly overlapping profiles for both pristine and doped $\alpha$-Fe$_2$O$_3$, indicating isotropic optical responses. Additionally, the EELS spectra show significant changes in peak intensity and spectral broadening resulting from doping with B and Y. These alterations suggest that doping affects the dielectric response and the collective electronic excitations of the material.

%
\begin{table}[htbp]
\centering
\caption{Summary of several optical properties, including optical band gap energy ($E_g$), static dielectric constant ($\epsilon_{\rm re,0}$), and the peak loss energy ($E_{\rm peak}$), of pristine and doped $\alpha$-Fe$_2$O$_3$.}
\resizebox{0.48\textwidth}{!}{%
\begin{tabular}{c c c c}
\Xhline{3\arrayrulewidth}
    Materials & $E_g$ & $\epsilon_{\rm re,0}$ & $E_{\rm peak}$   \\
     Name &  (eV) & -- & (eV)   \\
    \Xhline{2\arrayrulewidth}
    pristine  $\alpha$-Fe$_2$O$_3$  & $2.30$ & $7.809$ & $2.717$  \\
    B-doped $\alpha$-Fe$_2$O$_3$    & $1.65$ & $7.294$ & $2.762$  \\
    Y-doped $\alpha$-Fe$_2$O$_3$    & $2.25$ & $8.002$ & $2.672$  \\
    (B, Y)-doped $\alpha$-Fe$_2$O$_3$   & $1.58$ & $8.195$ & $2.678$  \\
    \Xhline{3\arrayrulewidth}
\end{tabular}}
\label{Table1}
\end{table}
%
%

Additionally, the calculated $E_g$ supports the analysis of the dielectric function and provides deeper insights into light-harvesting capability, optical response, and carrier transport characteristics. Pristine $\alpha$-Fe$_2$O$_3$ exhibits $E_g$ of 2.30 eV, corresponding to an absorption edge of approximately 540 nm. Y-doping slightly narrows $E_g$ to 2.25 eV, resulting in a modest red shift of the absorption edge while improving electronic transport and optical properties. In contrast, B doping significantly reduces $E_g$ to 1.65 eV, extending optical absorption into the red region and strengthening light-matter interactions through impurity-assisted transitions. Notably, (B, Y) co-doping reduces $E_g$ further to 1.58 eV while providing the most balanced dielectric behavior, a strong polarization response, and optimal optical losses. Consequently, the co-doped system combines broad-spectrum absorption with improved carrier transport, making it particularly attractive for applications requiring enhanced optical sensitivity, extended spectral coverage, and efficient photon-to-charge conversion. A summary of the key optical parameters, including $E_g$, $\epsilon_{\rm re,0}$, and the peak loss energy ($E_{\rm peak}$), for pristine and doped $\alpha$-Fe$_2$O$_3$ is presented in Table 1.

\subsubsection{Refractive Index and Extinction Coefficient and Electric Susceptibility}

The parameters $n$ and $k$ present complementary insights into the interaction of electromagnetic radiation with materials. The $n$ is related to electronic polarizability, which influences light propagation and confinement within a material \cite{born2013principles}. In contrast, $k$ measures the material's inherent ability to absorb and attenuate photons. An increase in $k$ indicates a higher electronic transition at specific photon energy levels. This aspect is typically associated with an increase in the joint density of states (JDOS) and optical selection rules that facilitate significant interband excitations. Therefore, the spectral behavior of $n$ and $k$ reflects changes in electronic structure, optical absorption processes, and carrier excitation dynamics, all of which affect the overall performance of optical materials.

%
\begin{figure}[htb]
    \centering
    \includegraphics[width =0.83\linewidth]{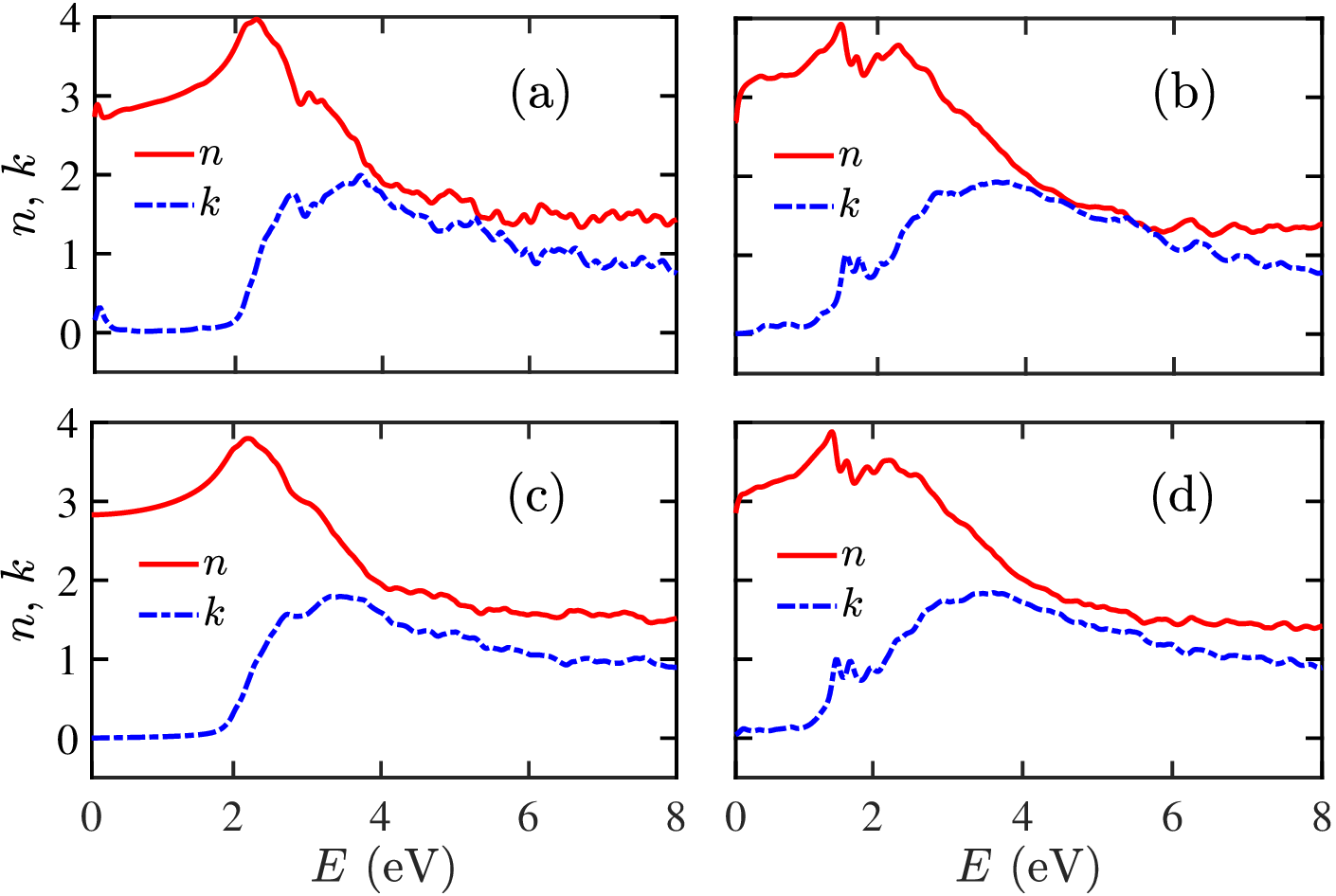}
    \caption{Refractive Index ($n$) and extinction Coefficient ($k$) as a function of energy ($E$) for (a) pristine $\alpha$-Fe$_2$O$_3$, (b) B-doped $\alpha$-Fe$_2$O$_3$, (c) Y-doped $\alpha$-Fe$_2$O$_3$, and (d) (B, Y)-co-doped $\alpha$-Fe$_2$O$_3$.}
    \label{Fig5}
\end{figure}
%
%

Figure 5 illustrates the characteristics of $n$ and $k$ for both pristine and doped $\alpha$-Fe$_2$O$_3$ varying with $E$ from zero to 8 eV. For pristine $\alpha$-Fe$_2$O$_3$, the peak of $n$ is observed in the near-visible range at $\sim$2 eV, overlapping with the initial absorption due to transitions between the 2p orbitals of O and the 3d orbitals of Fe, corresponding to the material's indirect band gap. Beyond $E=2$ eV, $n$ gradually decreases as refractive dispersion weakens outside the main absorption range. In contrast, $k$ increases sharply around $E=2$ eV, reaching a peak value between $E=3$ to $4$ eV, indicating strong interband excitations associated with deeper valence states. This optical behavior highlights relatively low to moderate absorption of solar light for pristine $\alpha$-Fe$_2$O$_3$.

B-doped $\alpha$-Fe$_2$O$_3$ shows significant modifications in both optical constants, as demonstrated in Fig.~5(b). The parameter $n$ remains high, reaching a maximum of about $3.9$, while $k$ exhibits a notable tail below $E=2$ eV and peaks at $E \approx 1.9$. These changes are attributed to B-induced localized states within the band gap, which create additional channels for optical transitions. Consequently, the spectral response extends toward lower $E$, enhancing light-matter interactions and optical absorption.

In comparison, Y-doped $\alpha$-Fe$_2$O$_3$ exhibits slightly lower peak values of $n$ and $k$, broader and smoother spectral features, as shown in Fig.~5(c). This change is likely due to the larger ionic radius of Y$^{3+}$ compared to Fe$^{3+}$, which affects the distortion of the Fe--O octahedra and alters the overlap of Fe--O orbitals. Such lattice strain modifies crystal-field splitting and may subtly reduce recombination rates. The result is a more homogeneous optical response with less scattering and improved carrier delocalization. Co-doping with (B, Y) combines the beneficial features of both dopants. The enhanced low-energy absorption introduced by B is maintained, while the spectra of $n$ and $k$ become smoother and less abrupt compared to pristine and B-doped $\alpha$-Fe$_2$O$_3$. Consequently, (B, Y) co-doping performs a favorable balance between broad spectral absorption, strong optical confinement, and efficient carrier transport.

%
\begin{figure}[htb]
    \centering
    \includegraphics[width =0.83\linewidth]{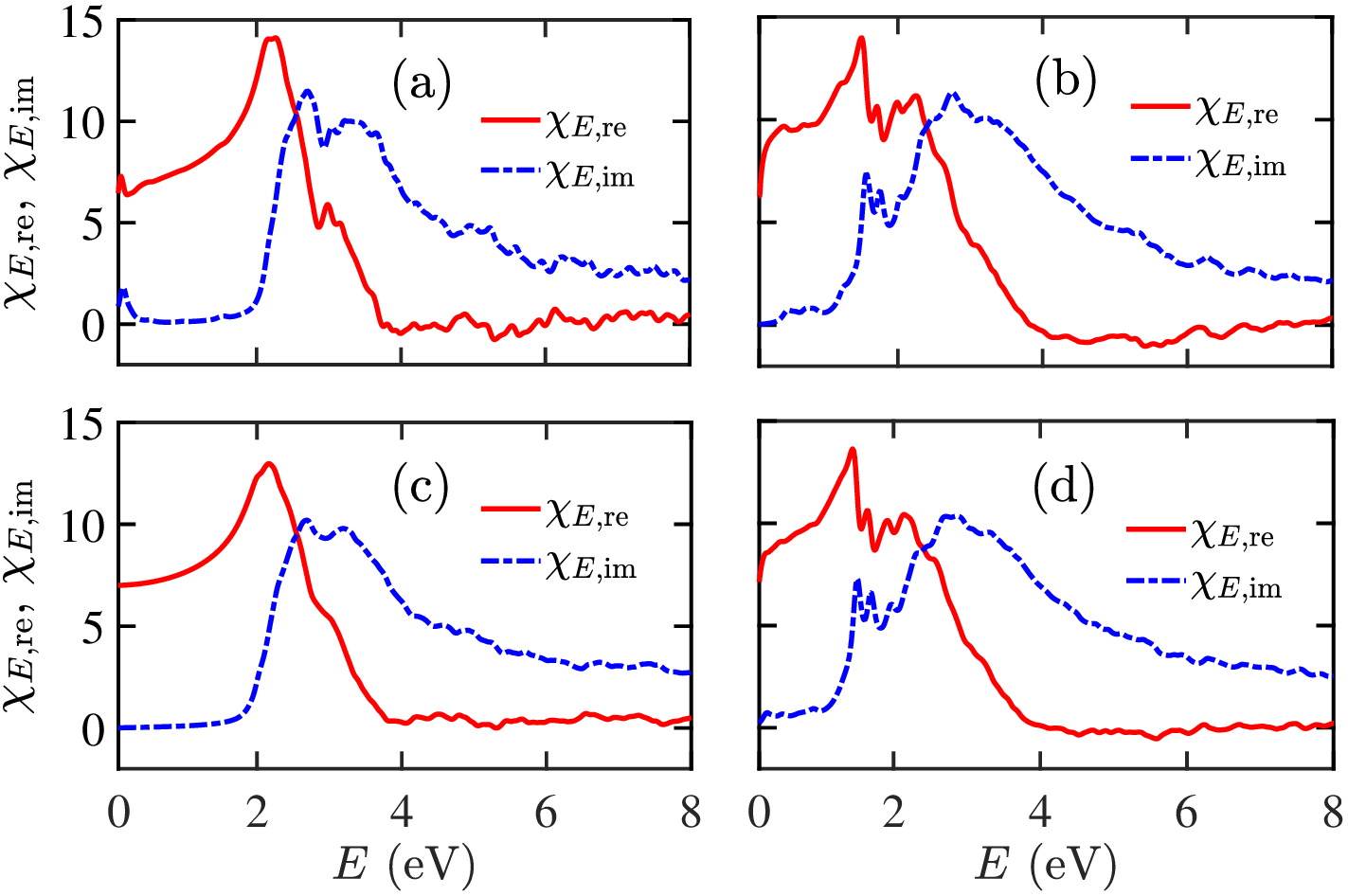}
    \caption{Real ($\chi_{E,{\rm re}}$) and imaginary parts ($\chi_{E,{\rm im}}$) of electric susceptibility ($\chi_E$) as a function of energy ($E$) for (a) pristine $\alpha$-Fe$_2$O$_3$, (b) B-doped $\alpha$-Fe$_2$O$_3$, (c) Y-doped $\alpha$-Fe$_2$O$_3$, and (d) (B, Y)-co-doped $\alpha$-Fe$_2$O$_3$.}
    \label{Fig6}
\end{figure}
%
%

The parameter $\chi_E$ represents a key material property that describes the electronic charge distribution within a material under an applied electric field. Its spectral behavior indicates the strength of electronic polarization, optical transitions, and charge-screening effects. Figure 6 illustrates the real ($\chi_{E,{\rm re}}$) and imaginary ($\chi_{E,{\rm im}}$) components of $\chi_E$ as a function of $E$ for both pristine and doped $\alpha$-Fe$_2$O$_3$. For pristine $\alpha$-Fe$_2$O$_3$, $\chi_{E,{\rm re}}$ exhibits a distinct peak around $E = 2$ eV, which corresponds to the fundamental absorption edge for transitions from O's 2p-orbitals to Fe's 3d-orbitals. This peak leads to an increase in $\chi_{E,{\rm im}}$, which reaches its maximum slightly above $E_g$. This behavior reveals the multiple interband transitions in this energy range. Beyond $E = 5$ eV, both components gradually decline, reflecting reduced transition probabilities in the deep ultraviolet range. The offset between the maxima of $\chi_{E,{\rm re}}$ and $\chi_{E,{\rm im}}$ is typical of resonant optical systems and results from the intrinsic coupling between dispersion and absorption.

In Fig.~6(b), B-doped $\alpha$-Fe$_2$O$_3$ shows a significant enhancement of $\chi_{E,{\rm im}}$ below $E = 2$ eV while maintaining a strong polarization response. This low-energy susceptibility arises from B-induced states, which introduce additional optical transition channels within the band gap and extend the spectral response toward lower photon energies. The broadening of the $\chi_E$ profile suggests that polarization can be sustained over a wider energy range, thereby enhancing light-matter interaction, while potentially increasing defect-assisted optical losses. In contrast, Y-doped $\alpha$-Fe$_2$O$_3$ exhibits a slightly reduced \(\chi_E\) with smoother spectral features, as shown in Fig.~6(c). This change indicates a redistribution of transition strength due to modified Fe--O hybridization. In the case of co-doping, (B, Y)-doped $\alpha$-Fe$_2$O$_3$ possesses the enhanced low-energy response introduced by B doping, while showing a more gradual susceptibility profile compared to pristine and B-doped $\alpha$-Fe$_2$O$_3$. This behavior suggests that Y moderates the localization effects associated with B-induced states, resulting in a broader and better uniform polarization response. Overall, the synergistic effects of (B, Y) co-doping provide the most balanced combination of broad optical excitation and moderated defect-related effects, making this material particularly attractive for applications that require efficient light-matter interaction and a stable electromagnetic response.

%
\begin{figure}[htb]
    \centering
    \includegraphics[width =0.89\linewidth]{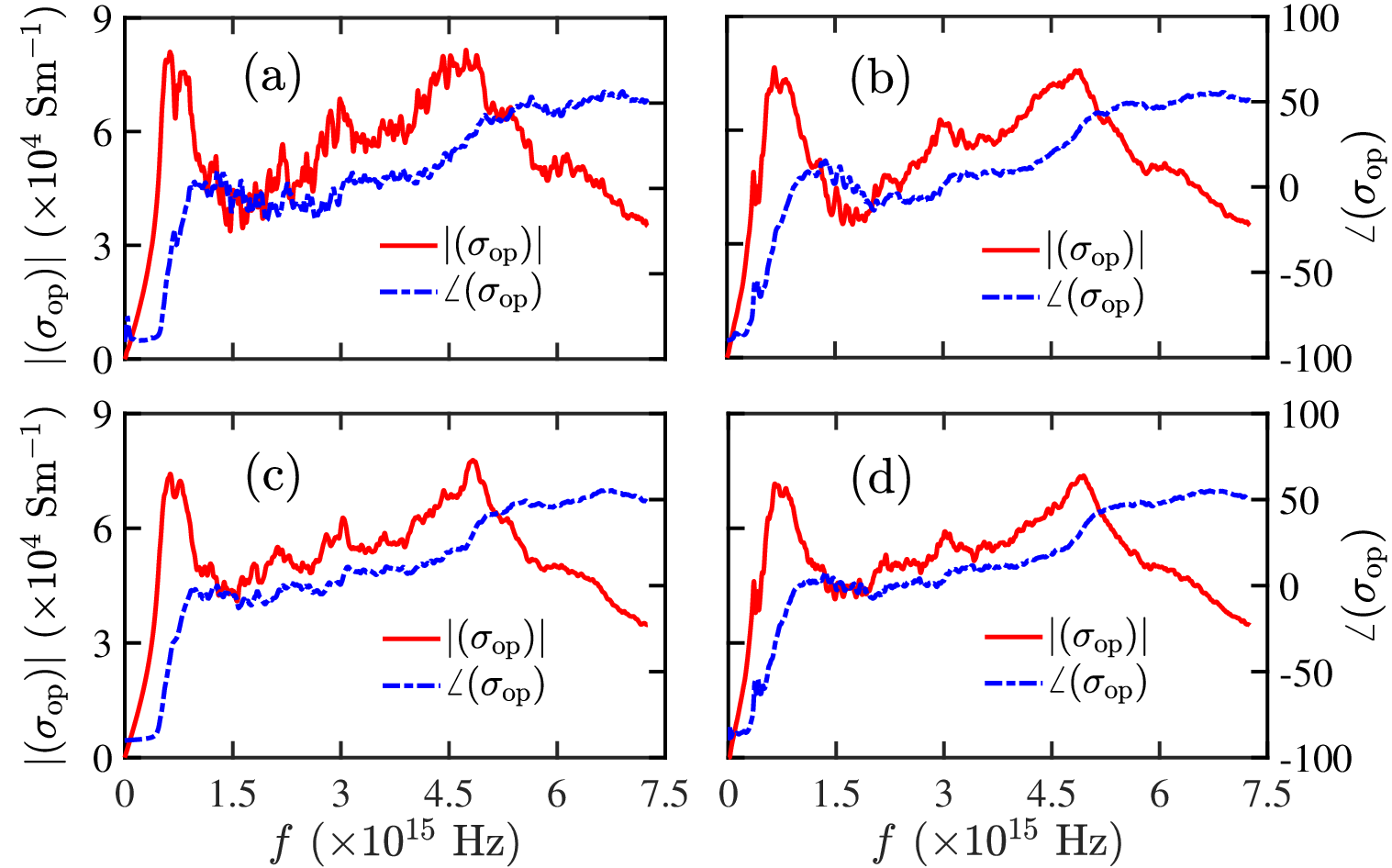}
    \caption{Absolute value, $|\sigma_{\rm op}|$, and polar angle, $\angle \sigma_{\rm op}$, in degree of optical complex conductivity, $\sigma_{\rm op}$, as a function of frequency, $f$, for (a) pristine pristine $\alpha$-Fe$_2$O$_3$, (b) B-doped $\alpha$-Fe$_2$O$_3$, (c) Y-doped $\alpha$-Fe$_2$O$_3$, and (d) (B, Y)-co-doped $\alpha$-Fe$_2$O$_3$.}
    \label{Fig7}
\end{figure}
%
%

%
\subsubsection{Optical Complex Conductivity}

The parameter $\sigma_{\rm op}$ measures the interaction between electromagnetic radiation and charge carriers in a material. It presents the efficiency of converting absorbed optical energy into electronic current. Consequently, the magnitude of the optical conductivity ($|\sigma_{\rm op}|$) is closely related to optical absorption and carrier excitation, while the phase angle ($\angle\sigma_{\rm op}$) reflects the balance between dissipative effects and polarization-driven responses. Figure 7 provides a comprehensive analysis of the $\sigma_{\rm op}$ for both pristine and doped $\alpha$-Fe$_2$O$_3$ as a function of frequency ($f$) in the optical regime. For pristine $\alpha$-Fe$_2$O$_3$, $|\sigma_{\rm op}|$ increases rapidly with $f$, exhibiting prominent maxima of approximately $7.5$--$8.0 \times 10^4$ S/m near $0.8 \times 10^{15}$ Hz and $4.5 \times 10^{15}$ Hz. These features align with strong interband electronic transitions previously identified in the dielectric and absorption spectra \cite{piccinin2019band}. The parameter $\angle \sigma_{\rm op}$ shows significant variations across this frequency range, indicating strong dispersion and complex interactions between electrons and photons.

B-doped $\alpha$-Fe$_2$O$_3$ demonstrates a broader range of conductivity while maintaining a similar maximum magnitude, as depicted in Fig.~7(b). The peaks in $|\sigma_{\rm op}|$ are broadened and intensified, indicating the introduction of defect states and structural distortions in the material. This broadening allows for a wider frequency range of conductivity responses rather than concentrating on specific resonances. While this improvement enhances spectral utilization and light-matter interactions, the defect-derived states may also lead to increased carrier localization and energy dissipation.

Y-doped $\alpha$-Fe$_2$O$_3$ exhibits slightly lower conductivity peaks, reaching approximately $7.0$--$7.5 \times 10^4$ S/m, along with a smoother phase evolution. The reduced sharpness of both $|\sigma_{\rm op}|$ and $\angle\sigma_{\rm op}$ suggests a more uniform distribution of optical transition probabilities due to modified Fe--O hybridization. Co-doping with B and Y further refines the optical conductivity characteristics. In (B, Y)-co-doped $\alpha$-Fe$_2$O$_3$, $|\sigma_{\rm op}|$ displays more distinct peaks and enhanced conductivity over a wider frequency range, indicating synergistic effects from the dopants that effectively tune the electronic structure. This behavior suggests a more continuous distribution of optically accessible states and a more stable carrier response under electromagnetic excitation. 

\subsubsection{Optical Reflectivity and Penetration Depth}

The parameters $R_{\text{op}}$ and $\delta_{\text{op}}$ describe the interaction of electromagnetic radiation with a material. The parameter $R_{\text{op}}$ measures the fraction of incident light that is reflected from the surface, while $\delta_{\text{op}}$ describes the distance over which optical intensity declines within the material \cite{fox2010optical}. Concurrently, these parameters influence photon injection, optical confinement, and the spatial distribution of absorbed energy, making them essential parameters of the optical performance of semiconductors.

%
\begin{figure}[htb]
    \centering
    \includegraphics[width =0.95\linewidth]{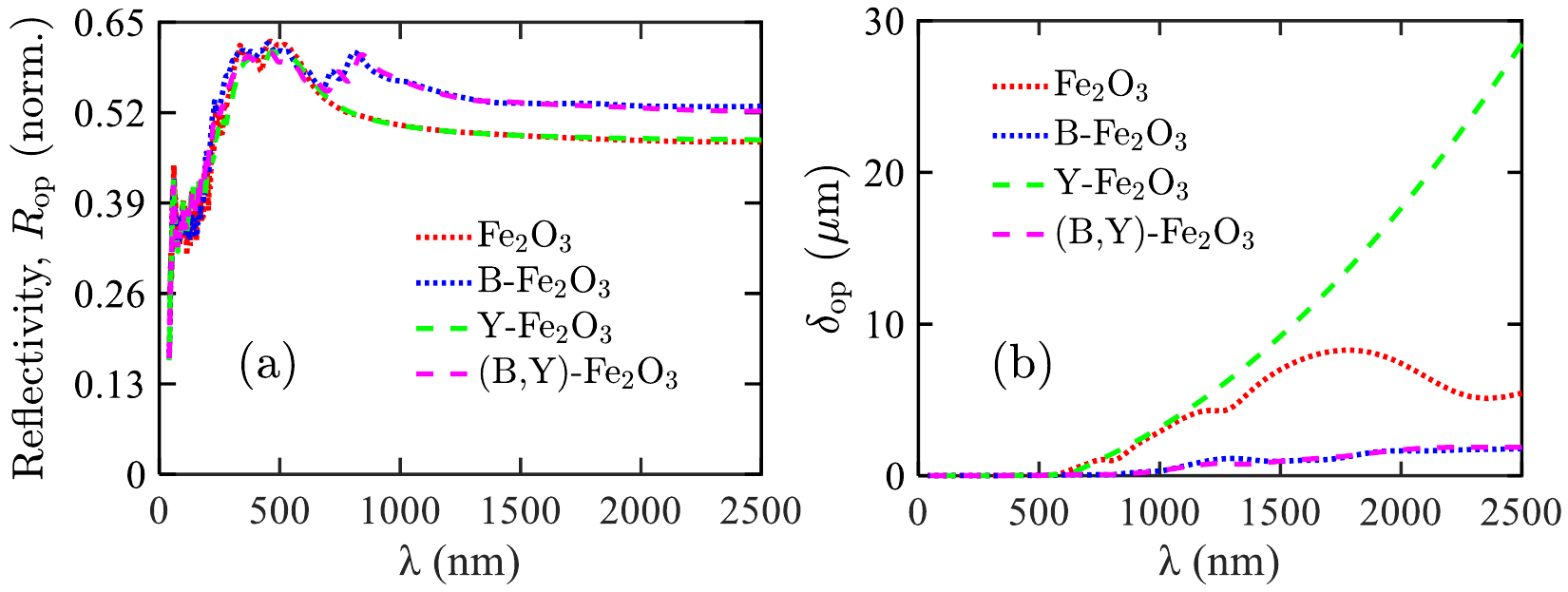}
    \caption{(a) Optical reflectivity, $R_{\rm op}$, and (b) penetration depth, $\delta_{\rm op}$, of pristine and doped $\alpha$-Fe$_2$O$_3$.}
    \label{Fig8}
\end{figure}
%
%

Figure 8(a) shows that pristine and doped $\alpha$-Fe$_2$O$_3$ exhibit similar reflectivity characteristics over the range of $\lambda $. In the ultraviolet region ($\lambda <  300$ nm), $R_{\text{op}}$ increases sharply from approximately 0.20 to 0.35, peaking at around 0.60 to 0.62 near $\lambda = 400-500$ nm. This high reflectivity is attributed to strong electronic polarization and the large refractive index near fundamental optical transitions. Beyond $\lambda < 600$ nm, $R_{\text{op}}$ gradually decreases and stabilizes between 0.48 and 0.53. The relatively minor differences among the pristine and doped systems suggest that doping has a limited effect on surface optical losses. However, B-doped and (B, Y)-co-doped $\alpha$-Fe$_2$O$_3$ demonstrate slightly higher reflectivity at longer wavelengths, indicating subtle modifications to the refractive response due to changes in the electronic structure from doping.

In contrast, the penetration depth shows a much stronger dependence on doping, as illustrated in Fig.~8(b). For pristine $\alpha$-Fe$_2$O$_3$, $\delta_{\text{op}}$ remains below 1 {\textmu}m for $\lambda < 800$ nm, confirming significant optical attenuation near the absorption edge. As $\lambda$ increases, $\delta_{\text{op}}$ steadily rises, reaching $\sim 8$ {\textmu}m near $\lambda = 1800$ nm before slightly decreasing to $\sim 5$ {\textmu}m at  $\lambda = 2500$ nm. This behavior reflects a reduction in interband absorption at lower photon energies, allowing light to penetrate deeper into the material. Y-doped $\alpha$-Fe$_2$O$_3$ shows the most substantial increase in penetration depth, reaching nearly 28 {\textmu}m at $\lambda = 2500$ nm. This significant enhancement indicates much weaker optical attenuation and a lower extinction coefficient in the long-wavelength region, allowing photons to travel considerably farther before being absorbed. Conversely, B-doped and (B, Y)-co-doped $\alpha$-Fe$_2$O$_3$ maintain relatively small penetration depths, remaining below $\sim 2$ {\textmu}m over most of the infrared region. Since $\delta_{\text{op}}$ is inversely related to the absorption coefficient, this reduced $\delta_{\text{op}}$ is consistent with the increased low-energy absorption observed in the extinction coefficient and optical conductivity spectra. Overall, these results highlight that doping primarily alters the spatial distribution of absorbed optical energy rather than the surface reflectivity, offering an effective approach to tailor $\alpha$-Fe$_2$O$_3$ for various photoactive, optoelectronic, and photonic applications.

%
%

%
%
\section{Conclusion} 

In summary, this study provides a comprehensive understanding of dopant-induced vibrational stability and optical response of $\alpha$-Fe$_2$O$_3$ through first-principles and Phonopy calculations. The analysis of phonon dispersion and optical properties indicates that both pristine and Y-doped $\alpha$-Fe$_2$O$_3$ systems maintain dynamic stability. However, doping with B leads to phonon instabilities due to lattice distortions in the Fe--O framework. Despite these structural instabilities in B-doped $\alpha$-Fe$_2$O$_3$, there is a significant improvement in optical absorption due to the formation of mid-gap states, enabling effective absorption in the range of approximately $E = 1.5$--$3.5$ eV. In contrast, Y doping enhances structural stability and moderates optical transitions by increasing orbital hybridization and reducing defect-related disruptions, resulting in a smoother dielectric and optical response. Co-doping with B and Y further enhances lattice stability and optical activity, as Y suppresses soft phonon modes and B introduces extra electronic states at the valence band that extend into the visible and near-infrared regions. This synergistic interaction makes (B, Y)-co-doped $\alpha$-Fe$_2$O$_3$ a well-balanced material for semiconductor and photonic technologies. Overall, the results underscore the potential of tailoring dopant chemistry to facilitate precise engineering of interband transition pathways and optical energy windows. This strategy transforms $\alpha$-Fe$_2$O$_3$ into a broadband, tunable photoactive material, providing a promising, stable, and efficient material for optoelectronic and solar energy conversion applications.

%
%

\section*{Supplementary Material}
The supplementary material presents the designed crystal structures of both pristine and doped $\alpha$-Fe$_2$O$_3$. The supplementary material also displays the optical tensor and electron energy loss spectroscopy (EELS) data along the $x$, $y$, and $z$ axes as a function of energy ($E$) for both pristine and doped $\alpha$-Fe$_2$O$_3$.  

\section*{Data Availability}
All data of the paper are presented in the main text and the supplementary material.

%
\section*{Author Declaration} 
The authors have no conflicts to disclose.

\section*{Acknowledgments} 
The authors sincerely appreciate the support and resources provided by the Photonics Laboratory and the Department of Electrical and Electronic Engineering at Bangladesh University of Engineering and Technology. Additionally, the authors declare that this work was carried out without any external financial support.

%

%
%
\small
\bibliographystyle{ieeetr}
\bibliography{references}

%
\end{document}